\documentclass[twocolumn,showpacs,preprintnumbers,amsmath,amssymb,10pt,aps,prx,superscriptaddress]{revtex4-1}
\usepackage{epsfig,amsopn}
\usepackage{graphicx}
\usepackage{epstopdf}
\usepackage{sidecap}
\usepackage{amsmath,amssymb,amsbsy}
\usepackage{amsthm}
\usepackage{enumerate}
\usepackage{float}
\usepackage{color}

\newcommand{\p}{\tilde{\Psi}}

\newcommand{\be}{\begin{equation}}
\newcommand{\ee}{\end{equation}}

\newcommand{\bea}{\begin{eqnarray}}
\newcommand{\eea}{\end{eqnarray}}

\newcommand{\Ket}[1]{{| #1 \rangle}}
\newcommand{\Bra}[1]{{\langle #1 |}}

\renewcommand{\vec}[1]{{\bf #1}}

\begin{document}
\title{Quantized large-bias current in the anomalous Floquet-Anderson insulator}
\author{Arijit Kundu}
\affiliation{Department of Physics, Indian Institute of Technology Kanpur, Kanpur 208016, India}
\affiliation{Physics Department, Technion, 320003, Haifa, Israel}
\author{Mark Rudner}
\affiliation{Niels Bohr International Academy and Center for Quantum Devices,
University of Copenhagen, 2100 Copenhagen, Denmark}
\author{Erez Berg}
\affiliation{Department of Condensed Matter Physics, Weizmann Institute of Science, Rehovot, Israel 76100}
\affiliation {Department of Physics, University of Chicago, Chicago, Illinois 60637, USA }
\author{Netanel H. Lindner}
\affiliation{Physics Department, Technion, 320003, Haifa, Israel}

\begin{abstract}
We study two-terminal transport through two-dimensional periodically driven systems in which all bulk Floquet eigenstates are localized by disorder.
We focus on the Anomalous Floquet-Anderson Insulator (AFAI) phase, a topologically-nontrivial phase within this class, which hosts topologically protected chiral edge modes 
coexisting with its fully localized bulk.  
We show that the unique properties of the AFAI yield remarkable far-from-equilibrium transport signatures: for a large bias between leads, a quantized amount of charge is transported through the system each driving period.
Upon increasing the bias, the chiral Floquet edge mode connecting source to drain becomes fully occupied and the current rapidly approaches its quantized value. 
\end{abstract}

\maketitle

Topological phenomena, such as the quantum Hall effect \cite{Klitzing1980} and Thouless' adiabatic pump \cite{ThoulessPump}, are characterized by the precise quantization of certain transport properties. 
Recently, periodic driving has emerged as a versatile tool to control the topological characteristics of quantum systems~\cite{Niu2007,Oka2009, KBRD, Inoue2010, Lindner2011, Jiang2011, 
Lindner2013, Gu11, Floquet_Transport_Kitagawa, Delplace2013, Podolsky2013,  ErhaiPRL, Iadecola2013, Goldman2014, Grushin2014, kundu2013, Kundu2014, Titum2015}.
Such ``Floquet'' systems can be realized in a wide variety of physical settings, including cold atomic, optical, and electronic systems~\cite{Wang_2013,Rechtsman_2013,Jotzu_2014,Desbuquois_2017}.
The extent to which Floquet systems may host quantized transport is an important direction of investigation.

Interestingly, periodically-driven quantum systems host unique topological phases which cannot be realized by their static counterparts~\cite{KBRD,WindingNumber, 1D_Majorana, Chiral_1D, 2D_TR, TopologicalSingularities, AFAI, KhemaniPRL2016, RoyHarper2016b, ElseNayakSPT, vonKeyserlingk2016a, Potter2016, RoyHarper2016a,
vonKeyserlingkPRB2016,
vonKeyserlingk2016b,
 NayakFloquetTimeCrystals, TimeCrystalObservation}.
The richer topological classification of these systems is due to their discrete (rather than continuous) time translation symmetry, which is manifested as a periodicity of the quasienergy -- the energylike variable that characterizes the Floquet spectrum. Crucially, this structure provides the basis for wholly new types of quantized transport phenomena, also without analogues in static systems. 

The first example of a quantized transport phenomenon unique to periodically-driven systems was uncovered in Ref.~\cite{ThoulessPump}.
There, Thouless showed that the charge transmitted through an insulating one-dimensional system is quantized as an integer multiple of the fundamental charge  when the system is adiabatically driven through a cycle in parameter space.

\begin{figure}[t]
\centering
\includegraphics[width=0.45\textwidth]{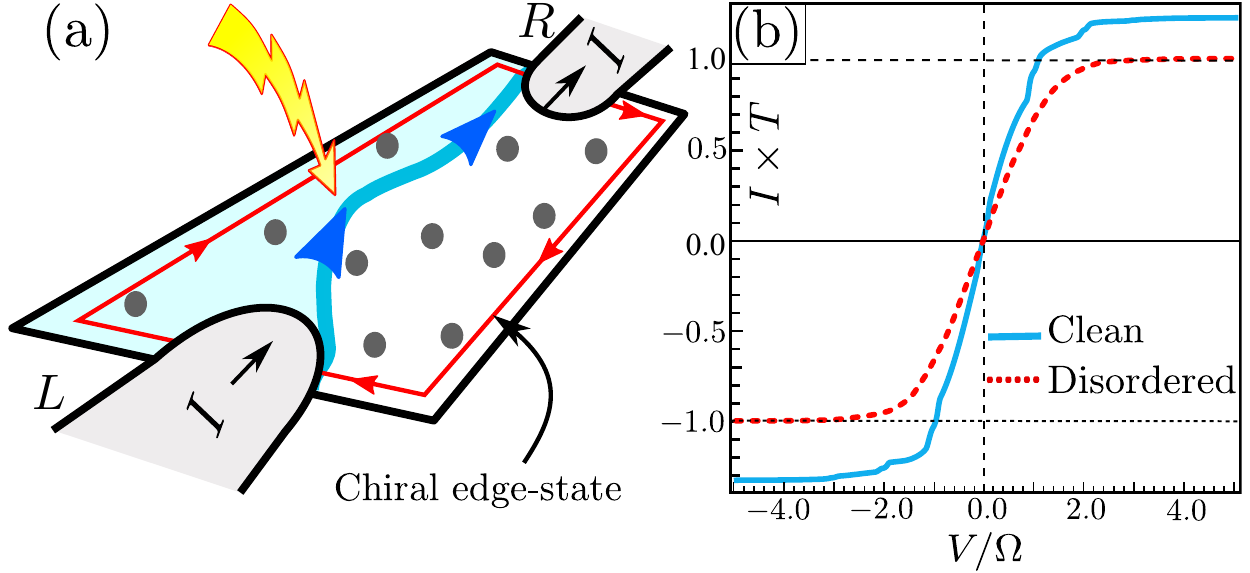}
\caption{Quantized transport in the AFAI phase. (a) Two-terminal transport setup. 
A large source-drain bias ensures that the edge states running from source to drain are fully filled, while those running from drain to source are empty.
(b) Bias ($V$) dependence of the steady state current, $I$, for clean (light blue) and fully-localized (dark red) systems.
The current saturates to the quantized value $I = 1/T$ for $V \gtrsim 2\Omega$, where $\Omega = 2\pi/T$ is the driving frequency.}
\label{fig:setup}
\end{figure}

More recently, in Ref.~\cite{AFAI} it was shown that two-dimensional, disordered,
periodically driven systems host a unique topological phase called the Anomalous Floquet Anderson
Insulator (AFAI). In the AFAI phase, all bulk Floquet eigenstates are localized, while chiral edge
states run along the system's boundaries. The AFAI's chiral edge states exist at {\it all} quasienergies; each such chiral edge mode carries
a quantized current when completely filled. In this work we show that, in a two-terminal transport
setup, the AFAI carries a \textit{net} quantized current $I = \mathcal{W}_{\rm 2D}/T$ in the limit
of large source-drain bias (see Fig.~\ref{fig:setup}). Here $\mathcal{W}_{\rm 2D}$ is the winding
number invariant that characterizes 2D periodically driven systems~\cite{WindingNumber, AFAI,
Magnetization}. Associated with the quantized current, we find an inhomogeneous density profile in
which the AFAI's right-moving chiral edge state is fully occupied, while the left-moving chiral
edge state is empty. Importantly, while quantized pumping in the Thouless pump is found in the
adiabatic limit, the large-bias quantized current carried by a driven system in the AFAI phase occurs for intermediate driving frequencies (comparable to the system's natural bandwidth).

The AFAI phase occurs in two-dimensional systems, whose dynamics are governed by a time-periodic Hamiltonian $H_S(t) = H_S(t + T)$, where $T$ is the driving period. The periodic driving gives rise to a unitary evolution $U_S(t) = \mathcal{T} e^{-i\int_0^t\!dt' H_S(t')}$, where $\mathcal{T}$ denotes time ordering.
The spectrum of the Floquet operator $U_S(T)$, given by $U_S(T) \Ket{\psi_n(0)} = e^{-i\varepsilon_nT}\Ket{\psi_n(0)}$, defines the Floquet states $\{\Ket{\psi_n(t)}\}$ and their quasienergies $\{\varepsilon_n\}$.

To study quantized transport in the AFAI phase, we consider a finite region of AFAI connected to two wide-bandwidth (non-driven) leads, as shown in Fig.~\ref{fig:setup}a.
The leads are indexed by $\lambda = \{{\rm L, R}\}$, standing for the left and right leads, respectively.
Dynamics of the combined system-lead setup are described by the Hamiltonian
\be
\label{eq:FullHam} H(t) = H_{\rm S}(t) + \sum_{\lambda = {\rm L,R}} H_{\lambda} + \sum_{\lambda = {\rm L,R}} H_{S\lambda},
\ee
where $H_{\rm S}(t) = H_{\rm S}(t + T)$ is the time-periodic Hamiltonian of the AFAI system, 
$H_{\lambda}$ is the Hamiltonian describing lead $\lambda$, and $H_{{\rm S}\lambda}$ describes the coupling between the system and lead $\lambda$.
We treat each lead as an ideal Fermi reservoir, with filling characterized by an equilibrium Fermi-Dirac distribution with chemical potential $\mu_\lambda$ in lead $\lambda$.
Specific forms for the Hamiltonian terms above will be given below. Throughout this paper, we use $e,\hbar=1$.

\begin{figure}[t]
\centering
\includegraphics[width=0.48\textwidth]{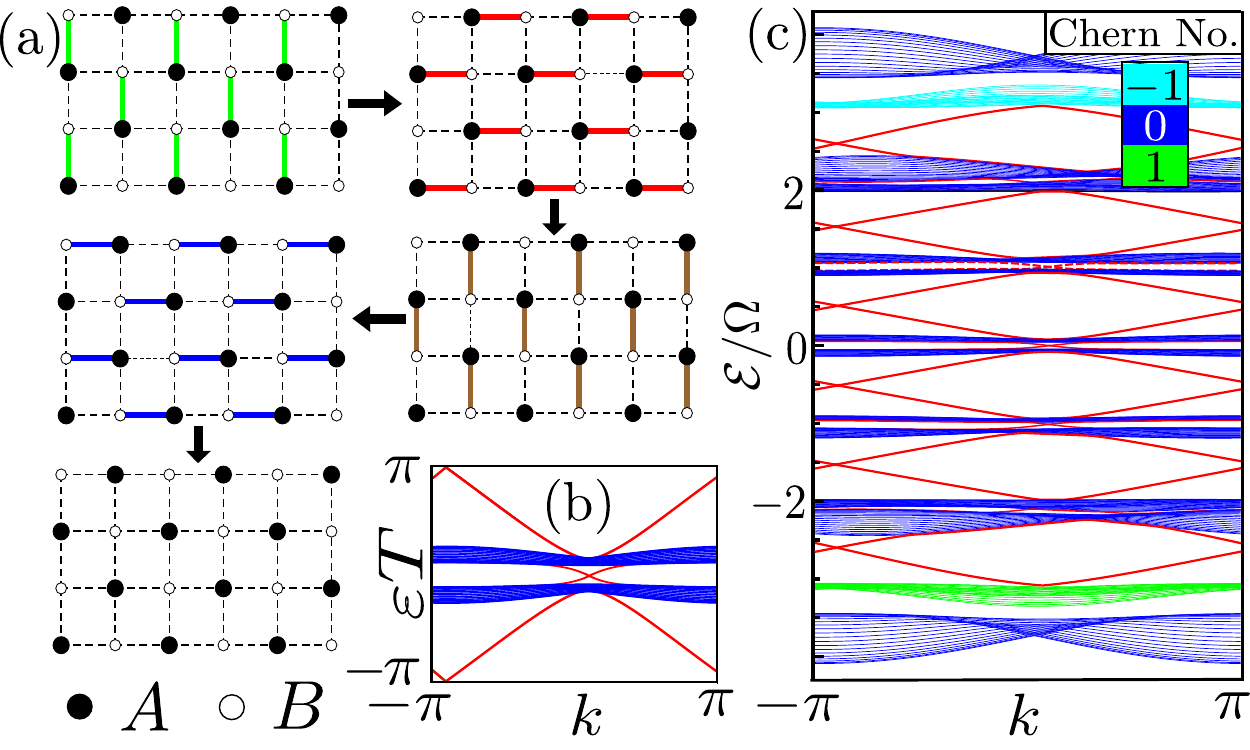}
\caption{Model of the AFAI phase. (a) Driving proceeds in 5 steps of equal length, $T/5$.
In each step, the highlighted bonds are active with strength $J_{ij} = 5\pi/(2T)$, Eq.~(\ref{eq:HamClean}), while all others are set to 0. The sublattices A and B are denoted by black and white circles, respectively.
(b) Quasienergy spectrum of $H_S^{\rm clean}$, with $D=\pi/(2T)$. 
(c) Spectrum of the truncated extended zone (EZ) picture system Hamiltonian, $\mathcal{H}^{\rm EZ}_S$ (with $M = 3$), see Eq.~(\ref{eq: extended ham}), in the absence of disorder.
While the bands near $\mathcal{E} = 0$ have Chern number zero, 
close to the truncation we find bands with Chern numbers $\pm 1$.
}
\label{fig:setup0}
\end{figure}

The AFAI phase can be realized by a variety of driving protocols and experimental platforms, including solid-state and cold atoms. For concreteness and simplicity, here we use the square lattice tight-binding model introduced in Ref.~\cite{AFAI}. In this model, the AFAI is described by the Hamiltonian $H_S(t) = H_S^{\rm clean}(t) + \sum_i w_i c^\dagger_i c_i$, where $c^\dagger_i$ ($c_i$) is the fermionic creation (annihilation) operator for site $i$, and $w_i$ is a normally-distributed on-site disorder potential with zero mean and standard deviation $w$.
The clean (disorder-free) Hamiltonian is given by 
\begin{align}\label{eq:HamClean}
 H_{\rm S}^{\rm clean}(t) = \sum_{\langle ij \rangle}J_{ij}(t)c^{\dagger}_ic_j + \sum_{i}D n_ic^{\dagger}_ic_i,
\end{align}
where $\{J_{ij}(t)\}$ are time-dependent nearest-neighbor hopping amplitudes. It is convenient to define two sublattices $A$ and $B$ on the square lattice (see Fig.~\ref{fig:setup0}). The piecewise-constant amplitudes $J_{ij}(t)$ connecting the two
sublattices are modulated according to the five step cycle depicted in Fig.~\ref{fig:setup0}a,
where each step has length $T/5$. Within each step, all nonzero hopping amplitudes (bold bonds)
have strength $J = \frac{5\pi}{2T}$; in the fifth interval, all $J_{ij}=0$.
The parameter $D$ is a staggered potential on the $A$ and $B$ sublattices, with $n_i= +1\ (-1)$ for the $A (B)$ sub-lattice.  We emphasize that the quantization of the current at large-bias is universal and independent of the specific model; a cold atom realization based on Refs.~\cite{MoriasSmith,Esslinger} is analyzed in the Appendix. 

Within the AFAI phase, realized for nonzero $w$ below a critical value~\cite{AFAI}, the system in an open geometry exhibits chiral edge sates in coexistence with a fully-localized bulk.
These chiral edge states are illustrated in the example spectra for the clean system ($w = 0$) in an infinite-strip geometry, shown in Fig.~\ref{fig:setup0}b.

We now study the steady-state current transported through the system when it is coupled to leads.
To this end, we consider the Heisenberg equations of motion for the operators $c_j(t)=U(t)c_j(t_0)U^\dagger(t)$ and $a^\lambda_j(t)=U(t)a_j(t_0)U^\dagger(t)$, where $a^\lambda_j$ is the fermionic annihilation operator on site $j$ of lead $\lambda$, and $U(t)=\mathcal{T}e^{-i\int_{t_0}^t dt' H(t')}$ is the evolution operator for the full Hamiltonian, Eq.~(\ref{eq:FullHam}).

To simplify notation we introduce the  operator vectors ${\bf a}_\lambda = (\cdots a_{\lambda i} \cdots)^T$ and ${\bf c} = (\cdots c_i \cdots)^T$, and express the system, lead, and system-lead coupling Hamiltonians in Eq.~(\ref{eq:FullHam}) as $H_S(t) = {\bf c}^\dagger {\bf H}_S(t)\, {\bf c}$, $H_\lambda={\bf a}^\dagger_\lambda {\bf H}_\lambda{\bf a}_\lambda$ and $H_{S\lambda} = {\bf c}^\dagger {\bf H}_{S\lambda} {\bf a}_\lambda + {\rm h.c.}$, respectively.
We leave the specific forms of the matrices ${\bf H}_\lambda$ and ${\bf H}_{S\lambda}$ unspecified for now.

The macroscopic leads are assumed to be attached in the very long past, such that the system operators ${\bf c}(t)$ are completely determined by the distribution in the leads; i.e., there is no memory of any initial occupations in the system. 
We then write a formal solution for the Heisenberg equation of motion, $i\dot{\bf c} =  {\bf H}_S\, {\bf c} + \sum_\lambda {\bf H}_{S\lambda}{\bf a}_\lambda$:
%
\begin{equation}
{\bf c}(t)=\int dt' \boldsymbol{G}(t,t') \left[\sum_{\lambda} {\bf H}_{S\lambda} {\bf g}_\lambda(t'-t_0){\bf a}_\lambda(t_0)\right],
\label{eq:c_solution}
\end{equation}
where ${\bf g}_\lambda(t)=-i\exp(-i{\bf H}_\lambda t)\theta(t)$ is the retarded propagator for lead $\lambda$ and $\boldsymbol{G}(t, t')$ is the full Green's function within the system. For the calculations below, it is convenient to furthermore define the Fourier-transformed Floquet Green's function,
\begin{align}
\boldsymbol{G}^{(k)}(\mathcal{E}) = \frac{1}{T}\int_0^Tdt\int_{-\infty}^\infty ds~\boldsymbol{G}(t,t-s)e^{i\mathcal{E} s}e^{ik\Omega t},
\label{eq: Green}
\end{align}
and
\begin{equation}
 \boldsymbol{\xi}_{\lambda}(\mathcal{E})={\bf H}_{S\lambda}\boldsymbol{\rho}_\lambda(\mathcal{E}){\bf H}^\dagger_{S\lambda},
 \label{eq: xi}
 \end{equation}
where $\boldsymbol{\rho}_\lambda(\mathcal{E}) = \sum_n \delta(\mathcal{E} - E_{\lambda n}) \Ket{\lambda n}\Bra{\lambda n}$ captures the density of states of lead $\lambda$, with ${\bf H}_\lambda \Ket{\lambda n} = E_{\lambda n}\Ket{\lambda n}$ \cite{footnoteone}.

The net current flowing into the right lead, averaged over one period, is given by
\begin{equation}
I= \frac{1 }{T}\int_0^T dt\, i\langle \left[H(t),N_{\rm R}(t)\right]\rangle, 
\label{eq: current def}
\end{equation}
where $N_{\rm R}(t)= {\bf a}_{\rm R}^\dagger(t) {\bf a}_{\rm R}(t)$ is the number operator for the right lead. 
Through Eq.~(\ref{eq:c_solution}) we express the {\it system} operators ${\bf c}(t)$ as linear combinations of the {\it lead} operators ${\bf a}_\lambda(t_0)$ in the distant past (we take $t_0 \rightarrow -\infty$).
Similarly, the lead operators ${\bf a}_\lambda(t)$ can be written in terms of ${\bf a}_\lambda(t_0)$.
We assume that the state in each lead $\lambda$ is given by a Fermi distribution $f_\lambda$ with chemical potential $\mu_\lambda$ and temperature $T_\lambda$: $\langle a_{\lambda n}^\dagger(t_0) a_{\lambda m}(t_0) \rangle = \delta_{nm} f_{\lambda}(\epsilon_{\lambda n})$, where $a^\dagger_{\lambda n}$ creates an electron in eigenstate $\Ket{\lambda n}$ in lead $\lambda$ (see above).
Using Eqs.~(\ref{eq:c_solution})-(\ref{eq: xi}) and the Fermi distributions for the leads, a standard calculation gives~\cite{Kohler}: 
%
\begin{eqnarray}
\label{eq:J_solution}
 &&\!\!\!\!I = 2\pi \int_{-\infty}^\infty d\mathcal{E}\, \sum_k \left\{T^{(k)}_{\rm RL}(\mathcal{E})f_{\rm L}(\mathcal{E}) - T^{(k)}_{\rm LR}(\mathcal{E})f_{\rm R}(\mathcal{E})  \right\},\\
\nonumber &&T^{(k)}_{\lambda\lambda'}(\mathcal{E}) = {\rm Tr}\left[\boldsymbol{G}^{(k)\dagger}(\mathcal{E})\boldsymbol{\xi}_\lambda(\mathcal{E} + k\Omega)\boldsymbol{G}^{(k)}(\mathcal{E})\boldsymbol{\xi}_{\lambda'}(\mathcal{E}) \right].
\label{eq: landauer}\end{eqnarray}
Here $T^{(k)}_{\lambda\lambda'}(\mathcal{E})$ 
is the probability for an electron at energy $\mathcal{E}$ to be transmitted from lead $\lambda'$ to lead $\lambda$, along with the absorption of $k$ photons from the driving field.


As we now show, the steady-state time-averaged current carried by the AFAI, Eq.~(\ref{eq:J_solution}), is {\it quantized} in the limit of large bias, $V \rightarrow \infty$, with $\mu_{\rm L} = V/2,\mu_{\rm R}\to -V/2$.
In this limit we may set $f_{\rm L}(\mathcal{E}) = 1$ and $f_{\rm R}(\mathcal{E}) = 0$, yielding
\begin{equation}
\label{eq:curr} I =\int_{-\infty}^{\infty} d\mathcal{E}\, \sigma(\mathcal{E}),\ \sigma(\mathcal{E}) = 2\pi \sum_k T^{(k)}_{\rm RL}(\mathcal{E}).
\end{equation}
\begin{figure}[t]
\centering
\includegraphics[width=0.47\textwidth]{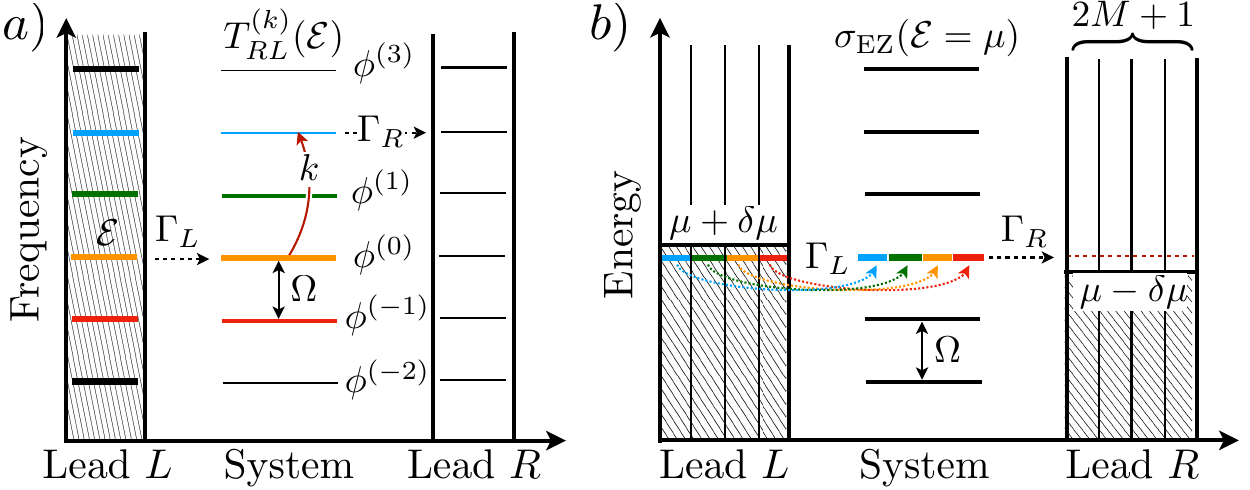}
\caption{Considerations leading to the sum rule in Eq.~(\ref{eq: sum rule}). (a) Transport through the driven system. A particle with energy $\mathcal{E}$ enters the system from the left lead via the component $\Ket{\phi^{(0)}_{\varepsilon}}$ of a Floquet state $\Ket{\psi_{\varepsilon}(t)}$, with quasi-energy $\varepsilon\approx\mathcal{E}$. The particle then scatters into a state with energy $\mathcal{E}+k\Omega$ in the right lead via its coupling to the component  $\Ket{\phi^{(k)}_{\varepsilon}}$. (b) Transport in the static extended zone (EZ) system, see Eq.~(\ref{eq: extended ham}). The EZ lead consists of $2M+1$ identical channels, shifted in energy by integer multiples of $\Omega$. A state in the lead with energy $\mu$ and harmonic index $n$ is coupled to the component $\Ket{\Phi^{(n)}_{\mu}}$ of the eigenstate $\Ket{\psi_\mu^{\rm EZ}}$ of $\mathcal{H}_S^{\rm EZ}$ with eigenvalue $\mu$.}\label{fig:sumrule}
\end{figure}

In the following, we show the quantization of the current by relating $\sigma(\mathcal{E})$ to the differential conductance of an associated {\it static} system.
For illustration, we first consider the dominant processes contributing to $\sigma(\mathcal{E})$, 
see Fig.~\ref{fig:sumrule}a. 
In each process a particle in the left lead with energy $\mathcal{E}$ scatters into a Floquet state of the system with quasienergy $\varepsilon \approx \mathcal{E} +n \Omega$~\cite{footnote:gamma}. 
The integer $n$ is determined by our convention for Floquet states, 
$\Ket{\psi_\varepsilon(t)}=e^{-i\varepsilon t}\sum_{m}\Ket{\phi_{\varepsilon}^{(m)}} e^{-i\Omega m t}$, with $-\Omega/2\leq \varepsilon<\Omega/2$.
The scattering process thus proceeds through the coupling between the lead state and the component $\Ket{\phi_{\varepsilon}^{(-n)}}$.
The particle then scatters into a state in right lead with energy  $\mathcal{E}+ k\Omega$, via its coupling to the component $\Ket{\phi_{\varepsilon}^{(k-n)}}$.
Thus, in the process of scattering from the left to the right lead the particle absorbs $k$ photons from the time-periodic drive.
The collection of processes involving such changes in the particle's energy is captured by the sum appearing in the definition of $\sigma(\mathcal{E})$, Eqs.~(\ref{eq:J_solution}) and~(\ref{eq:curr}).

We now re-express the current, Eq.~(\ref{eq:curr}), as 
$I=\int_{-\Omega/2}^{\Omega/2} (dI/d\varepsilon) d\varepsilon$, with $dI/d\varepsilon=\sum_{n}\sigma(\mathcal{E}+n\Omega)$.
The quantity $dI/d\varepsilon$ can be related to the differential conductance of a \textit{static} system, which describes the periodically driven system in an ``extended zone'' (EZ)
frequency-space picture. The Hamiltonian of the static EZ system is given by $\mathcal{H}^{\rm EZ}=\sum^M_{m,n}H^{\rm EZ}_{mn}|m\rangle\langle n|$, where the sum runs over  $-M \le n,m \le M$, and
\begin{align}
 H^{\rm EZ}_{mn} =& -\delta_{mn}n\Omega+\int_0^{T}\frac{dt}{T}e^{i(m-n)\Omega t}H(t).
 \label{eq: extended ham}
\end{align}
The operator $\mathcal{H}^{\rm EZ}$ acts in enlarged Hilbert space, which is a tensor product of the original Hilbert space and a $(2M+1)$-dimensional auxiliary space, which we call the harmonic space.

 As in Eq.~(\ref{eq:FullHam}), we write $\mathcal{H}^{\rm EZ}=\mathcal{H}_S^{\rm EZ}+\sum_\lambda\mathcal{H}_{S\lambda}^{\rm EZ}+\sum_\lambda \mathcal{H}_{\lambda}^{\rm EZ}$.
An eigenstate of $\mathcal{H}_S^{\rm EZ}$ with energy $\mathcal{E}$
 can be expanded as $\Ket{\psi^{\rm EZ}_\mathcal{E}}=\sum_{n} \Ket{\Phi_{\mathcal{E}}^{(n)}}\otimes |n\rangle$.
The eigenvalues of $\mathcal{H}_S^{\rm EZ}$ in the range $-\Omega/2\! \leq \mathcal{E}\! < \Omega/2$ approximate the {\it quasienergy} spectrum of $U_S(T)=\mathcal{T}e^{-i\int_0^T dt H_S(t)}$,
becoming 
exact for 
 $M\to\infty$.
 Importantly, in this limit, for each $\Ket{\psi^{\rm EZ}_\mathcal{E}}$ there is a corresponding partner Floquet state with quasienergy $\varepsilon= \mathcal{E}+m\Omega$ (with $|\varepsilon| \le \Omega/2$) in the original driven problem: $\Ket{\psi_\varepsilon(t)}=e^{-i\varepsilon t}\sum_{n}\Ket{\phi_{\varepsilon}^{(n)}} e^{-i\Omega n t}$, with 
$\Ket{\phi_{\varepsilon}^{(n-m)}}=\Ket{\Phi_{\mathcal{E}}^{(n)}}$.

We now relate the relevant transport processes in the static EZ and Floquet pictures (see Fig.~\ref{fig:sumrule}). 
Consider the differential conductance, $\sigma_{\rm EZ}(\mu)$, of the EZ system described by ${\mathcal{H}}^{\rm EZ}$.
 Since the lead is not driven, the spectrum of $\mathcal{H}_{\lambda}^{\rm EZ}$  consists of $2M+1$ copies of that of $H_\lambda$, shifted by integer multiples of $\Omega$; it can thus be viewed 
as a lead with many channels, labeled by the harmonic index.
We define $\sigma_{\rm EZ}(\mu)$ by taking the Fermi level of the left and the right EZ leads to be $\mu+\delta\mu$ and $\mu-\delta\mu$, and take $-\Omega/2\leq \mu<\Omega/2$ throughout~\cite{footnotetwo}.

Consider now the dominant processes contributing to $\sigma_{\rm EZ}(\mu)$.
The system-lead coupling $\mathcal{H}_{S\lambda}^{\rm EZ}$ conserves the harmonic index.
Therefore, a lead state with energy $\mathcal{E}$ and harmonic index $n$ (which corresponds to a state of the physical lead with energy $\mathcal{E}-n\Omega$) is coupled to the state $\Ket{\psi^{\rm EZ}_\mathcal{E}}$ through the component $\Ket{\Phi_{\mathcal{E}}^{(n)}}$.
To obtain  $\sigma_{\rm EZ}(\mu)$, we sum the contributions of states with energies close to $\mu$ from all harmonic-index channels in both leads. 
Using the correspondence between $\{\Ket{\psi^{\rm EZ}_\mu}\}$ and 
$\{\Ket{\psi_\mu(t)}\}$, for $M \gg 1$, we thus obtain (for details, see Appendix):
\begin{equation}
  \sum_n\sigma(\mu+n\Omega) = \sigma_{\rm EZ} (\mu).
  \label{eq: sum rule}
 \end{equation}

Importantly, in the EZ picture, $\sigma_{\rm EZ}(\mu)$ is just the two-terminal differential conductance of a disordered Chern insulator, with $\mu$ lying in a mobility gap.
To see why this is the case, consider the spectrum of $\mathcal{H}^{\rm EZ}_S$ in the AFAI phase. In the spectral range $-\Omega/2\leq\mu<\Omega/2$ it exhibits two important properties: (i) all bulk states  are localized~\cite{footnote:delocal}, and (ii) chiral edge states exist at all energies within this range. These two properties of $\mathcal{H}^{\rm EZ}_S$ are a direct consequence of the properties of  $U_S(T)$ in the AFAI phase.
Since in the EZ picture the number of edge states corresponds to the total Chern number of all bulk states 
below $\mu$, 
the spectrum of $\mathcal{H}^{\rm EZ}_S$ must contain a band with nontrivial Chern number at an energy 
near the harmonic space truncation at $n = -M$. 
The quantized two-terminal differential conductance of such a Chern insulator~\cite{footnotethree}, $\sigma^{\rm EZ}(\mu) = \mathcal{W}_{\rm 2D}$,
together with Eq.~(\ref{eq:curr}), yields $I=\sum_n \int_{-\Omega/2}^{\Omega/2} d\mathcal{E} \sigma(\mathcal{E}+n\Omega)=\mathcal{W}_{\rm 2D}/T$.

For the model given in Fig.~\ref{fig:setup0}a, the above considerations are exemplified by inspecting the spectrum of the corresponding $\mathcal{H}^{\rm EZ}_S$ (without disorder),  given in Fig.~\ref{fig:setup0}c.
Here we find a single chiral edge in the spectral range $-\Omega/2\leq\mathcal{E}<\Omega/2$; in this spectral range, the Chern numbers of the bands are all zero.
However, the highlighted bands near the bottom and top of the spectrum, which are strongly affected by the truncation, have Chern numbers $\pm 1$.

\begin{figure}[tb]
\centering
\includegraphics[width=0.45\textwidth]{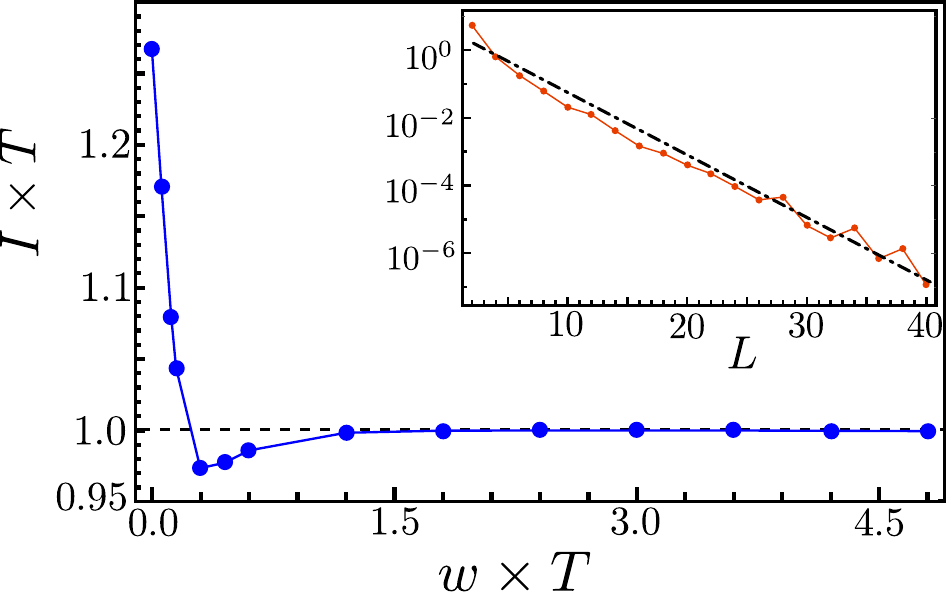}
\caption{Steady state current vs.~disorder strength $w$, for $V\gg\Omega$. As $w$ is increased from zero, the steady state current (averaged over a period) rapidly approaches the quantized value of $1/T$ \cite{footnote:transition}. The sample has dimensions $L\times W=40\times 20$ sites.
The leads are taken to have widths $W_0 = W/2$.
{\it Inset}: Bulk contribution to the steady state current, computed using a cylindrical geometry with contacts on opposite edges of the cylinder, for $w = 4.5/T$.
Exponential decay of the bulk contribution with increasing $L$ indicates that the system is in the localized regime.}
\label{fig:withDis}
\end{figure}

\begin{figure}
\centering
\includegraphics[width=0.45\textwidth]{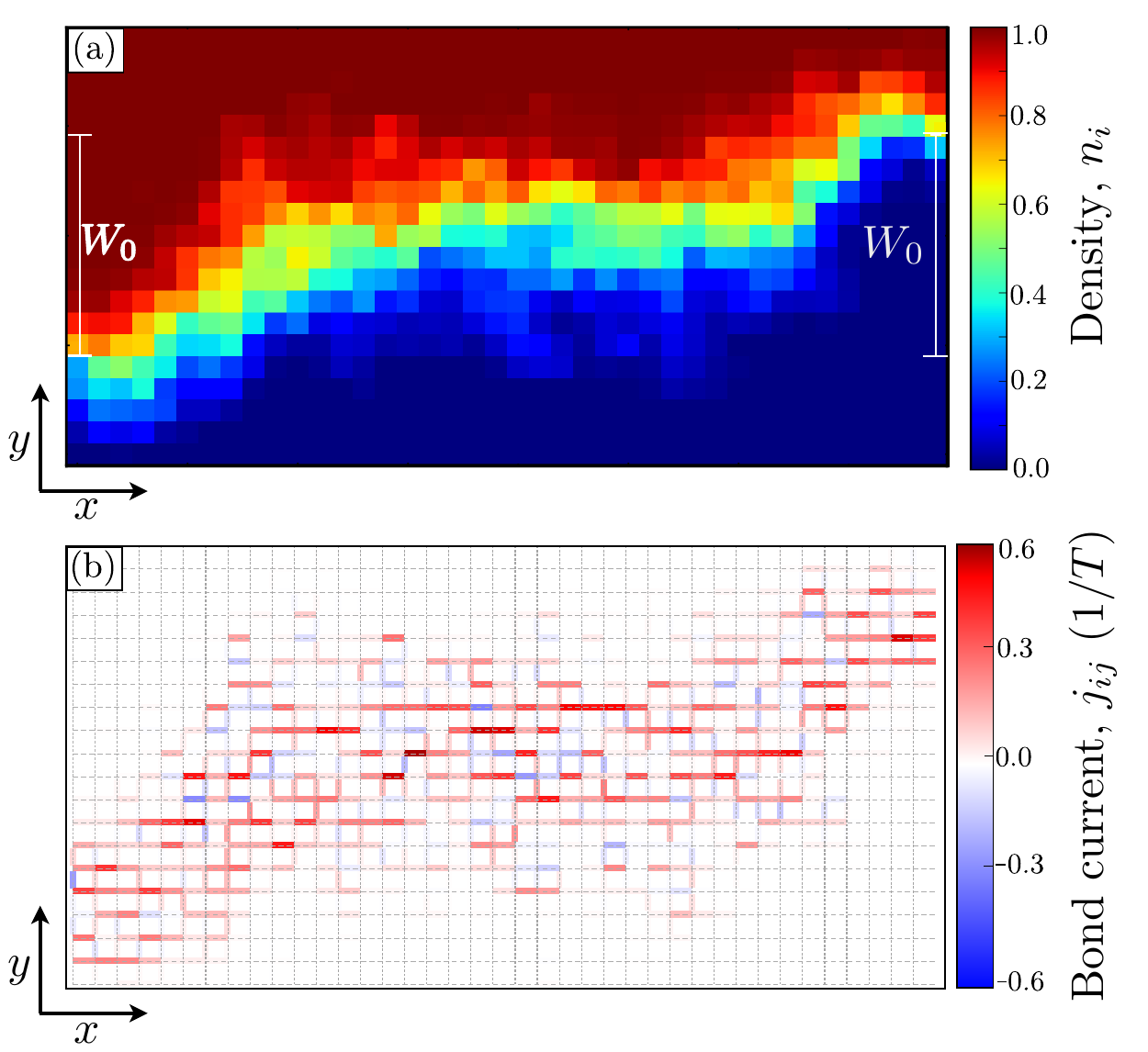}
\caption{(a) Map of the steady-state period-averaged density, $n_i$, for $w = 4.5/T$.
The large bias between the leads, $V \gg \Omega$, ensures that the edge state running from source to drain is fully occupied, while that running from drain to source is empty.
(b) The period-averaged bond currents $j_{ij}$ (see text). 
The current density is concentrated 
at the interface between fully occupied and empty regions.} 
\label{fig:CurDes}
\end{figure}

\emph{Numerical simulations.}---%
To support the arguments above, we now numerically study the steady state current.
We simulate the model described above, Eq.~(\ref{eq:FullHam}), for a range of system sizes and disorder strengths $w$, see Fig.~\ref{fig:withDis}. We take $D = \pi/(2T)$, and
 the leads to have 
constant density of states, $\rho_{0\lambda} = 1/J$.
The lead-system coupling ${\bf H}_{S\lambda}$ is taken to yield $\xi_\lambda(\mathcal{E})=\sum_{{\bf r}\in W_0}\rho_{0\lambda}|{\bf r}\rangle\langle {\bf r}|$, where the sum  runs over $W_0$ system sites directly adjacent to lead $\lambda$ (see Fig.~\ref{fig:CurDes}a).

In the presence of disorder, all bulk states are localized and the current through the bulk vanishes exponentially with the distance between the leads.
To probe this, we computed the current in a cylindrical geometry, 
with leads attached at opposite ends of the cylinder such that there were no edge states connecting the source and drain (shown in the inset of Fig.~\ref{fig:withDis}).
As shown in Fig.~\ref{fig:setup}b and the main panel of Fig.~\ref{fig:withDis}, for the Hall bar geometry of Fig.~\ref{fig:setup}a the total current through the system saturates to the quantized value $I = 1/T$ in the insulating regime, for large (finite) bias. 

As explained above, a quantized current is expected to flow in the AFAI when 
  the edge-states exiting from the left lead are completely filled, while those exiting the right lead are empty.
We confirm this picture (for a typical disorder realization) by mapping out the steady state time-averaged local density, $n_i = \frac{1}{T}\int_0^T dt\, \langle c^\dagger_i(t) c_i(t)\rangle$, in Fig.~\ref{fig:CurDes}a.
This situation is realized for ``good'' contacts, with appropriately strong couplings $\xi_\lambda$ and large enough contact width $W_0$ (see Fig.~\ref{fig:CurDes} and the Appendix). 

To further investigate the spatial distribution of the current, we map out the period-averaged bond current density, $j_{ij} = \frac{1}{T}\int_0^T dt\,2\text{Im}[J_{ij}(t)\langle c_{i}^{\dagger}(t)c_{j}(t) \rangle]$, see Eq.~(\ref{eq:HamClean}).
As shown in Fig.~\ref{fig:CurDes}b, the current density is concentrated at the boundary of the  filled and empty regions.
This result may at first seem counterintuitive since a) all states in the bulk are localized and b) we expect the quantized current to be carried by the chiral edge states.
However, it is crucial to remember that the local current density $j_{ij}(t)$ includes contributions of both transport currents and magnetization current~\cite{Magnetization, NiuRMP}.
The quantized {\it transport current} is indeed carried by the chiral edge states, as they are the only delocalized states in the system (also see the Appendix).

\emph{Summary.}---%
In this work we demonstrated theoretically a new topological quantized transport phenomenon, occurring in disordered two-dimensional periodically driven systems.
In contrast to 
the equilibrium quantized Hall {\it conductivity}, in the AFAI phase, which occurs in a system \textit{far from equilibrium}, we find a quantized {\it current} in the limit of large bias \cite{footnote:comparison}. Looking ahead, disorder-induced localization may provide a route for stabilizing interacting Floquet phases of matter by suppressing energy absorption from the periodic drive.
Recently, several works proposed interacting analogues of the AFAI \cite{Potter1,Potter2,Potter3,Potter4,Harper}.
Determining whether quantized transport and other response functions can be used to probe these interacting phases will be crucial for further progress in the field.

\emph{Acknowledgements.}---%
A. K. acknowledges the support from the Indian Institute of Technology - Kanpur and was supported in part at the Technion by a fellowship of the Israel Council for Higher Education. N. L. acknowledges support from the People Programme (Marie Curie Actions) of the European Union\textquoteright s Seventh Framework Programme (No. FP7/2007\textendash 2013) under REA Grant
Agreement No. 631696, from the Israeli Center of Research Excellence (I-CORE) \textquotedblleft Circle of Light.\textquotedblright{}. N. L. and E. B. acknowledge support from the European Research Council (ERC) under the European Union Horizon 2020 Research and Innovation Programme (Grant Agreement No. 639172). M. R. gratefully acknowledges the support of the European Research Council (ERC) under the European Union Horizon 2020 Research and Innovation Programme (Grant Agreement No. 678862), and the Villum Foundation. M. R. and E. B. acknowledge support from CRC 183 of the Deutsche Forschungsgemeinschaft.

\bibliographystyle{apsrev}

\begin{thebibliography}{50}
\bibitem{footnoteone} In the $\{\lambda n\}$ basis, we have $[\rho_\lambda(\mathcal{E})]_{nn} = \frac{-1}{2\pi i}{\rm Im}\left\{[g_\lambda(\mathcal{E})]_{nn}\right\}$.
\bibitem{footnote:gamma}
Due to finite coupling strength, the energy conservation condition is broadened on the scale of the tunneling rate.
\bibitem{footnotetwo} The states of  $\mathcal{H}_L^{\rm EZ}$  are populated according to a Fermi-Dirac distribution with chemical potential  $\mu+\delta \mu$ regardless of their harmonic index (and similarly for the right lead).
\bibitem{footnote:transition} The negative overshoot at low disorder is due to the aspect ratio of the system, see the Appendix. Note that at very strong disorder, the system undergoes a transition to a trivial insulator, where the current vanishes. This transition was studied in Ref.~\cite{AFAI}.
\bibitem{footnote:delocal} Note that delocalized states exist at energies close to the truncation. However, for sufficiently large M their energies do not lie in the window $–\Omega/2<\mu<\Omega/2$, and therefore do not contribute to transport.
\bibitem{footnotethree} The quantized value is obtained for sufficiently large contact length.
\bibitem{footnote:comparison} The I-V characteristic of quantum Hall systems is generically non-linear at large source-drain biases, but the current does not saturate to a universal quantized value as it does for the AFAI.
\end{thebibliography}

\appendix
\section{Sum rule}\label{app:SumRule}
In this section, we derive the sum rule appearing in Eq.~(10) of the main text.
Consider $|\phi_{\alpha}^{(k)\pm}\rangle$, the $k$th Fourier component of the time-periodic Floquet state $|\phi_{\alpha}^{\pm}(t)\rangle$, which satisfies
\begin{align}\label{eq:HamL}
\left({\bf H}_S(t) \mp i \boldsymbol{\Gamma} - i\frac{d}{dt}\right)|\phi_{\alpha}^{\mp}(t)\rangle = (\epsilon_{\alpha} \mp i\gamma_{\alpha})|\phi_{\alpha}^{\mp}(t)\rangle.
\end{align}
In the above $\boldsymbol{\Gamma} = \sum_\lambda \boldsymbol{\xi}_\lambda$, 
where we have assumed a constant density of states in each lead. The $\mp$ symbol in $|\phi_{\alpha}^{\mp}(t)\rangle$ labels the right and left eigenstates of the non-Hermitian operator $\left({\bf H}_S(t) - i \boldsymbol{\Gamma} - i\frac{d}{dt}\right)$, respectively. 
The Floquet states satisfy the completeness and orthogonality relations,
\begin{align}
\sum_{\alpha}|\phi^{-}_{\alpha}(t)\rangle\langle \phi^{+}_{\alpha}(t)| = \mathbb{I}, ~~ \int_{0}^{T}\frac{dt}{T}\langle \phi_{\alpha}^{-}(t)|\phi_{\beta}^{+}(t)\rangle=\delta_{\alpha\beta}.\nonumber
\end{align}
The Green's function satisfies
\begin{align}\label{eq:GF}
\left( i\frac{d}{dt} + \mathcal{E} - {\bf H}_S(t) + i \boldsymbol{\Gamma} \right)\boldsymbol{G}(t,\mathcal{E})=\mathbb{I},
\end{align}
with $\boldsymbol{G}(t,\mathcal{E}) = \sum_{k}e^{-ik\Omega t}\boldsymbol{G}^{(k)}(\mathcal{E})$.
The Floquet Green's function can thus be written as: 
\begin{align}\label{eq:GFF}
\boldsymbol{G}^{(k)}(\mathcal{E}) &=\sum_{\substack{\alpha \\ n\in\mathbb{Z}}}\frac{ |\phi_{\alpha}^{(n+k)-}\rangle\langle \phi^{(n)+}_{\alpha}| }{\mathcal{E} - \epsilon_{\alpha} -n\Omega + i\gamma_{\alpha}}.
\end{align}
As written in Eq.~(8) of the main text, the current is given by:
\begin{align}
I &=  2\pi  \sum_k \int_{-\infty}^{\infty} \text{d}\mathcal{E}~\text{Tr}\left[\boldsymbol{G}^{(k)\dagger}(\mathcal{E})\boldsymbol{\xi}_R~\boldsymbol{G}^{(k)}(\mathcal{E})\boldsymbol{\xi}_L\right],
\end{align}
and we define 
\begin{align}
\sigma(\mathcal{E}) = 2\pi  \sum_k \text{Tr}\left[\boldsymbol{G}^{(k)\dagger}(\mathcal{E})\boldsymbol{\xi}_R~\boldsymbol{G}^{(k)}(\mathcal{E})\boldsymbol{\xi}_L\right].
\end{align}
With this definition, the current is $I = \int_{-\infty}^{\infty}d\mathcal{E}~\sigma(\mathcal{E})$, or equivalently:
\begin{align}
\label{eq:sigmadIdE}I = \int_{-\Omega/2}^{\Omega/2}d\varepsilon~dI/d\varepsilon, \quad dI/d\varepsilon = \sum_{n\in\mathbb{Z}}\sigma(\varepsilon+n\Omega).
\end{align}
From Eq.~(\ref{eq:GFF}), we obtain:
\begin{align}
dI/d\varepsilon &= 2\pi\sum_{n,k} \text{Tr} \left[\boldsymbol{G}^{(k)\dagger}(\varepsilon+n\Omega)\boldsymbol{\xi}_{\rm R}\boldsymbol{G}^{(k)}(\varepsilon+n\Omega)\boldsymbol{\xi}_{\rm L}\right] \nonumber\\
& = 2\pi\sum\limits_{\substack{n,k \\ \alpha\beta \\ m,q}} \frac{ \langle \phi^{(m)+}_{\alpha}|\boldsymbol{\xi}_{\rm L}|\phi_{\beta}^{(q)+}\rangle }{\varepsilon - \epsilon_{\alpha} + (n-m)\Omega + i\gamma_{\alpha}}\times\nonumber\\
\label{eq:dIdE2} &\quad \quad \quad \quad \quad \times\frac{ \langle \phi^{(q+k)-}_{\beta}|\boldsymbol{\xi}_{\rm R}|\phi_{\alpha}^{(m+k)-}\rangle }{\varepsilon - \epsilon_{\beta} +(n-q)\Omega - i\gamma_{\beta}}.
\end{align}

Recall that our goal is to relate $dI/d\varepsilon$ to the differential conductance of the static extended-zone (EZ) system, and thereby derive Eq.~(10) of the main text.
The EZ Hamiltonian, $\mathcal{H}^{\rm EZ}$, is defined in the main text, Eq.~(9). Considering that the system-lead coupling is time independent, and assuming a constant density of states, 
i.e, $({H}_{S\lambda}^{EZ})_{mn} = \delta_{mn}H_{S\lambda}$, the EZ eigenstates satisfy:
\begin{equation}
\left(\boldsymbol{\mathcal{H}}_S \mp i \boldsymbol{\Gamma}^{\rm EZ} - i\frac{d}{dt}\right)|\Phi_{\alpha,p}^{\mp}\rangle = (\mathcal{E}_{\alpha,p} \mp i\gamma^{\rm EZ}_{\alpha,p})|\Phi_{\alpha,p}^{\mp}\rangle,
\end{equation}
with $\boldsymbol{\Gamma}^{\rm EZ}_{mn} = \delta_{mn}\boldsymbol{\Gamma}$, and where $-M\leq p \leq M$.
Together, $\alpha$ and $\p$ provide a complete labeling of the EZ eigenstates.
The left and right eigenstates form a complete set, with the relations: $\sum_{\alpha,p}|\Phi^{-}_{\alpha,p}\rangle\langle \Phi^{+}_{\alpha,p}| = \mathbb{I}, ~~ \langle \Phi_{\alpha,p}^{-}|\Phi_{\beta,p'}^{+}\rangle=\delta_{\alpha\beta}\delta_{pp'}$.
Correspondingly, the Green's function of the EZ system satisfies:
\begin{align}\label{eq:GFE}
\left( \mathcal{E} - \boldsymbol{\mathcal{H}}_S^{\rm EZ} + i \boldsymbol{\Gamma}^{\rm EZ} \right)\boldsymbol{\mathcal{G}}^{\rm EZ}(\mathcal{E})=\mathbb{I}.
\end{align}
We represent $\boldsymbol{\mathcal{G}}^{\rm EZ}(\mathcal{E})$ in the basis of eigenstates above as
\begin{align}
\label{eq:GEZ}\boldsymbol{\mathcal{G}}_{\rm EZ}(\mathcal{E}) &=\sum_{\alpha,p}\frac{ |\Phi_{\alpha,p}^{-}\rangle\langle \Phi^{+}_{\alpha,p}| }{\mathcal{E} - \mathcal{E}_{\alpha,p} + i\gamma^{\rm EZ}_{\alpha,p}}.
\end{align}

\begin{figure*}
	\includegraphics[width=0.9\textwidth]{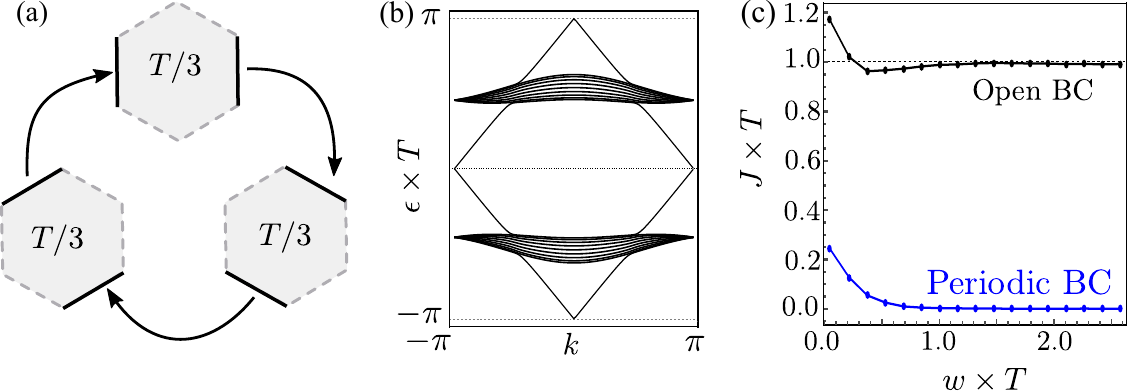}
	\caption{(a) Three step driving protocol for the non-disordered version of the AFAI. For each time step of duration $T/3$, only hoppings in a direction parallel to one of the three mirror plane directions of the hexagonal lattice are nonzero, and have strength $J>0$. b) The quasienergy spectrum for $\Omega =2\pi/T \approx 1.88J$ in a ribbon geometry shows two nearly flat bands and chiral edge states. An AFAI can be realized by introducing static on-site disorder. (c) In a two terminal setup at large bias, a quantized time-averaged current is obtained in the Hall bar geometry (open boundary conditions in the y direction), for sufficiently strong disorder strength $w$ such that the localization length is larger than system size. In a cylindrical geometry (closed periodic boundary conditions in the y direction), the current decays exponentially with increasing disorder strength.}\label{fig:3T}
\end{figure*}

The differential conductance of the EZ system at energy $\mu$ is given by
\begin{align}
\label{eq:SigmaEZ}\sigma_{\rm EZ}(\mu) = \frac{dI_{\rm EZ}(\mu)}{d\mu} = 2\pi \text{Tr}\left[\left.\boldsymbol{\mathcal{G}}^{\rm EZ}\right.^{\dagger}\!(\mu)\boldsymbol{\xi}^{\rm EZ}_R~\boldsymbol{\mathcal{G}}_{\rm EZ}(\mu)\boldsymbol{\xi}^{\rm EZ}_L\right],
\end{align}
where $(\boldsymbol{\xi}^{{\rm EZ}}_{\lambda})_{mn} = \delta_{mn}\boldsymbol{\xi}_\lambda$.
Substituting Eq.~(\ref{eq:GEZ}) into Eq.~(\ref{eq:SigmaEZ}), and writing $\Ket{\Phi^\pm_{\alpha,p}} = \sum_n \Ket{\Phi_{\alpha,p}^{(n)\pm}} \otimes \Ket{n}$, gives
\begin{align}
\sigma_{\rm EZ}(\mu) &= 2\pi \sum\limits_{\substack{\alpha,p,\beta,p'\\m,n}} \frac{ \langle \Phi^{(m)+}_{\alpha,p}|\boldsymbol{\xi}_{\rm L}|\Phi_{\beta,p'}^{(m)+}\rangle }{\mu - \mathcal{E}_{\alpha,p} + i\gamma^{\rm EZ}_{\alpha,p}}\frac{ \langle \Phi^{(n)-}_{\beta,p'}|\boldsymbol{\xi}_{\rm R}|\Phi_{\alpha,p}^{(n)-}\rangle }{\mu - \mathcal{E}_{\beta,p'} - i\gamma^{\rm EZ}_{\beta,p'}}.
\end{align}

We focus on value of $\mu$ between $-\Omega/2$ and $\Omega/2$.
Crucially, for $\mu$ in this range, the main contribution to $\sigma_{\rm EZ}(\mu)$ comes from states with $|\mathcal{E}_{\alpha,p} - \mu|\lesssim \gamma_{\alpha,p}$.
For $M$ [the parameter controlling the truncation of the EZ Hamiltonian, see main text around Eq.~(9)] much greater than 1, we have $\lim_{M\to\infty} \Ket{\Phi^{(m)\pm}_{\alpha,p}} = \Ket{\phi^{(m+p)\pm}_\alpha}$; the corresponding eigenvalues also satisfy $\lim_{M\to\infty} \mathcal{E}_{\alpha,p}  = \epsilon_\alpha+p\Omega$ and $\lim_{M\to\infty} \gamma^{\rm EZ}_{\alpha,p}  = \gamma_\alpha$.
Thus,
\begin{widetext}
	\begin{align}
	\sigma_{\rm EZ}(\mu) &= 2\pi\sum\limits_{\substack{\alpha,\beta\\m,n,p,p'}} \frac{ \langle \phi^{(m+p)+}_{\alpha}|\boldsymbol{\xi}_{\rm L}|\phi_{\beta}^{(m+p')+}\rangle }{\mu - \epsilon_{\alpha}-p\Omega + i\gamma_{\alpha}}\frac{ \langle \phi^{(n+p')-}_{\beta}|\boldsymbol{\xi}_{\rm R}|\phi_{\alpha}^{(n+p)-}\rangle }{\mu - \epsilon_{\beta} - p'\Omega - i\gamma_{\beta}}.
	\end{align}
\end{widetext}
After a relabeling of indices, the expression above matches that in Eq.~(\ref{eq:dIdE2}).
Thus, comparing with Eq.~(\ref{eq:sigmadIdE}), we have shown $\sum_n \sigma(\mu + n\Omega) = \sigma_{\rm EZ} (\mu)$.

\section{Cold Atom Realization}
\label{app:3step}
In this section we demonstrate the quantized two terminal transport in a model which is amenable to direct realization in a cold atom system. The model relies on a recent work   \cite{MoriasSmith},  which  considers a driven honeycomb optical  lattice for cold atoms. Ref.~\cite{MoriasSmith} proposes a driving protocol in which the driving period $T$ is divided into three time steps of length $T/3$. In each step the lattice is shaken with a high frequency along one of the three mirror plane directions of the honeycomb lattice. With appropriately chosen driving amplitude, the shaking can fully suppress the static hopping along bonds with components parallel to the shaking direction, while leaving the perpendicular bonds unaffected. By choosing the driving period $T$ to obey  $JT/3\approx\pi/2$, where $J$ is the hopping amplitude (for the unsuppressed bonds perpendicular to the shaking direction), an anomalous Floquet band structure with chiral edge states and zero bulk Chern numbers is obtained.

Building on that proposal, we consider a tight binding model on a honeycomb lattice subjected to a three-step piecewise constant driving protocol, see Fig.~\ref{fig:3T}a. In each step of the driving period, the only non-zero hopping amplitudes are those parallel to one of the three mirror planes. In Fig.~\ref{fig:3T}b, we show the quasienergy  band-structure for this model in a cylinder geometry (periodic boundary conditions in one direction), where the driving period $T$ and the hopping strength $J$ are related via $\Omega=2\pi/T \approx 1.88J $. The band structure features two nearly flat bands and chiral edge states along the edges of the cylinder. Introducing static onsite disorder realizes an AFAI phase. We numerically calculate the current averaged over a driving period in a two terminal setup, see Eq.~(6) in the main text. The results are plotted in  Fig.~\ref{fig:3T}c.  For the cylinder geometry, the current decreases exponentially with increasing disorder, indicating that the system is indeed in the (localized) AFAI phase. When computed in the Hall bar geometry, the system exhibits a quantized current for disorder strength large enough such that the localization length is smaller than the system size.




\section{Dependence on contact width}
\label{app:Width}
\begin{figure}
	\includegraphics[width=0.45\textwidth]{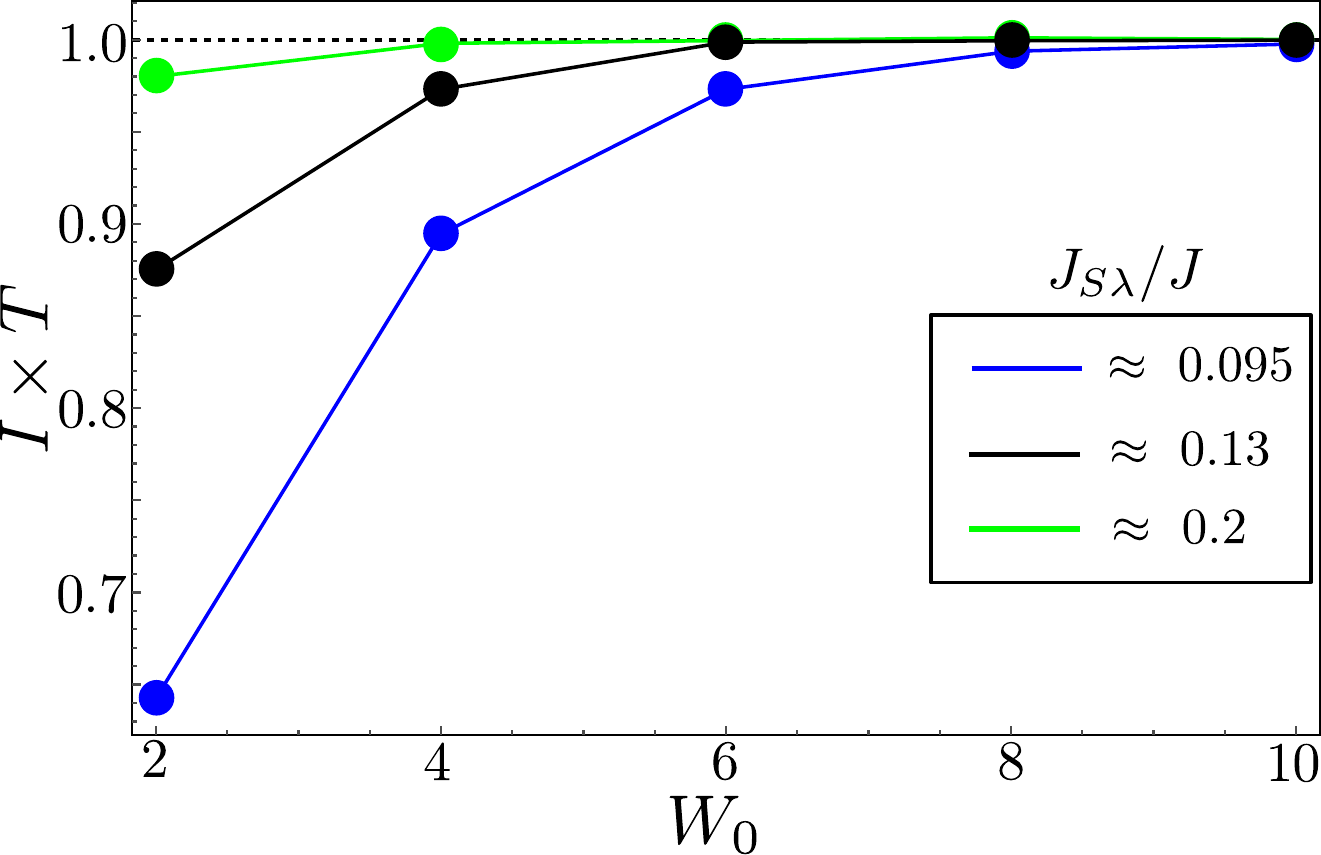}
	\caption{Convergence of current to the quantized value as a function of contact width $W_0$, for three values of the system-lead coupling $J_{S\lambda}$. 
		Here we used $L\times W= 40 \times 20$ sites, disorder strength $w\times T=1.5$, and constant density of states of the leads, $\rho_\lambda = 1/J$.}\label{fig:W0}
\end{figure}
\begin{figure}[t!]
	\includegraphics[width=0.45\textwidth]{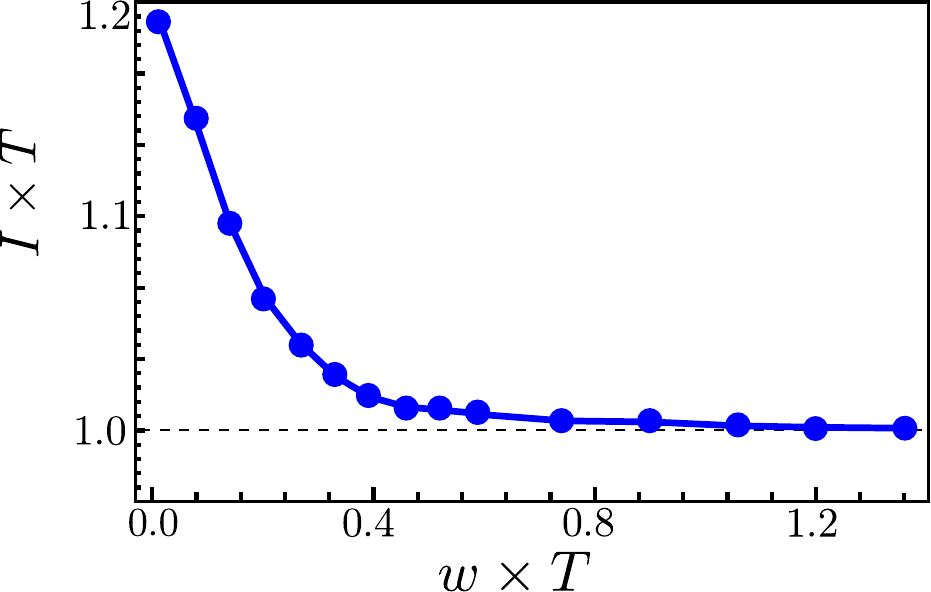}
	\caption{Steady state current vs.~disorder strength $w$, for $V\gg\Omega$, for a ``short and wide'' AFAI. As $w$ is increased from zero, the steady state current (averaged over a period) rapidly approaches the quantized value of $1/T$ \textit{from above}. The sample has dimensions $L\times W=20\times 30$ sites.
		The leads are taken to have widths $W_0 = W/2$.}\label{fig:aspect}
\end{figure}
In order to achieve quantized two-terminal transport, it is necessary that the chiral edge states fully equilibrate with the leads in the contact region.
For example, consider the current impinging on the drain contact.
It is essential that each particle reaching the drain is absorbed into the drain lead with unit probability, to avoid it returning back to the source and diminishing the net transmitted current.
The probability for the particle to be absorbed into the lead approaches one, exponentially with the width $W_0$ of the contact; the lengthscale associated with this exponential is controlled by the matrix elements $J_{S\lambda}$ governing hopping between the system and the lead. 

In Fig.~\ref{fig:W0} we demonstrate the convergence of the transmitted current as a function of $W_0$ and $J_{S\lambda}$.
We take $J_{S\lambda}$ to be constant for all sites within the contact region.
Here we focus on the weak coupling regime, where $J_{S\lambda}$ is small compared with the hopping matrix elements within the system, $J$.
As expected in this limit, the absorption length decreases with increasing $J_{S\lambda}/J$.

\section{Dependence on aspect ratio}
In Fig. 4 of the main text, we studied the steady state current averaged over a period, as a function of disorder strength $w$. For low values of $w$, the localization length of the bulk Floquet states is larger than the system size. In this regime, a transport current between the two leads can be carried by the bulk states. Therefore, for low values of $w$, we expect the current to be larger than the quantized value of $1/T$. This expectation is indeed confirmed by our numerical simulation, as shown in Fig 4.
In addition, Fig.~4 shows an interesting feature: as $w$ is increased further, the deviation from the quantized value of $1/T$ becomes \textit{negative}, and with increasing values of $w$, the current approaches the quantized value from below. This feature can be understood as follows: since the system studied numerically in Fig.~4 is ``long and narrow'' ($L=40, W=20$, where the leads are attached to the narrow edges), for intermediate values of disorder the localization length is smaller than the length of the system but larger than its width. Therefore, the bulk cannot transport particles between the leads, and at the same time there is still significant backscattering between the counterpropagating edge states on opposite sides of the sample. To confirm this explanation, we numerically simulated the transport experiment in a system which is ``wide and short'' ($L=20,W=30$, where the leads are attached to the wide edges). The results, shown in Fig.~\ref{fig:aspect}, indeed do not show any negative overshoot.

\section{Magnetization and transport current}
\label{app:current}
\begin{figure}[t!]
	\includegraphics[width=0.45\textwidth]{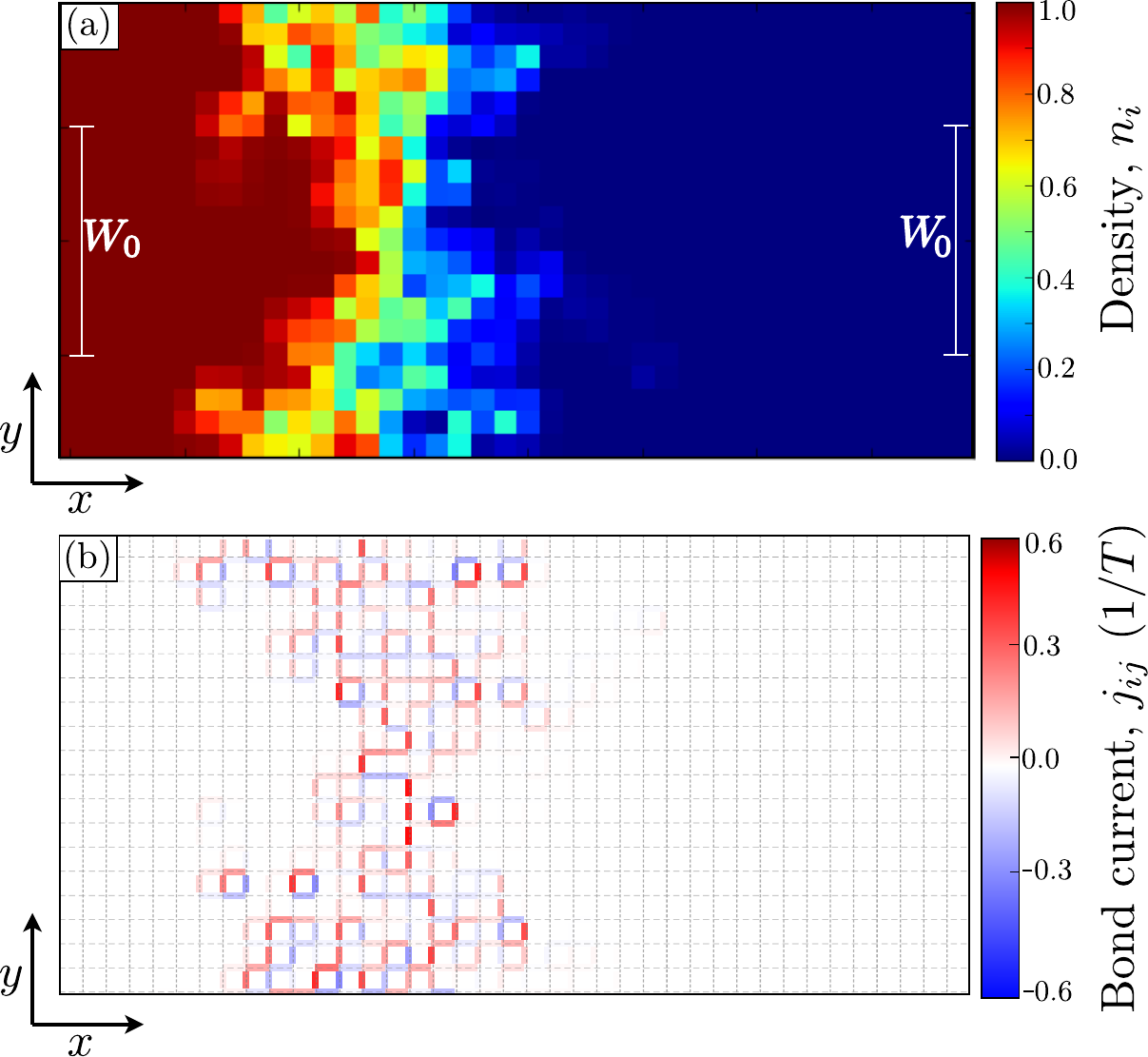}
	\caption{Density (a) and bond current (b) maps for an AFAI in the cylindrical geometry. The system has periodic boundary conditions along the y direction, and is connected to leads at x=0 and $x=L_x$, with Fermi levels $\frac{+V}{2}$ and $\frac{-V}{2}$. A single disorder realization is used.}\label{fig:current}
\end{figure}
The bond current is the sum of the magnetization current and the transport current.  The bulk of the AFAI is fully localized and is therefore insulating and cannot carry any transport current. It can, however, host magnetization currents in regions where the magnetization is inhomogeneous (recall that in a steady state the magnetization current is defined as ${\vec{j}}_m=\mathrm{\nabla}\times\vec{m})$.  Note that due to the aforementioned relation, steady-state magnetization currents (which are by nature circulating currents) do not contribute to the total current transported across any cut all the way through the system, and therefore cannot contribute to transport between the leads. For the AFAI in a Hall bar geometry, the only delocalized states that permit particles to propagate from one lead to the other are the AFAI’s topological edge states. Therefore any transport current between source and drain must be carried by the edge states.

Since the AFAI bulk cannot carry transport current, the density profile in a \textit{cylindrical geometry} should only exhibit a non-vanishing density near the lead with the larger Fermi energy. Likewise, the bond current can only be significant in a small region close to the lead where the particles are located (far away sites, being empty, cannot host currents). To numerically confirm this expected behavior, we computed the density and bond current profiles in the steady state in the cylindrical geometry. The results, shown in Fig.~\ref{fig:current}, directly confirm the above expectations demonstrating that the bulk does not carry any transport current between the two leads.

\end{document}